\documentclass{article}
\usepackage{amsfonts}
\usepackage{amsmath}
\usepackage{epsfig}
\usepackage{pstricks}
\usepackage{pst-grad} 

\def\beq{\begin{equation}}
\def\eeq{\end{equation}}
\def\bea{\begin{eqnarray}}
\def\eea{\end{eqnarray}}
\def\bq{\begin{quote}}
\def\eq{\end{quote}}
\def\simlt{\stackrel{<}{{}_\sim}}
\def\simgt{\stackrel{>}{{}_\sim}}

\begin{document}

\title{\hspace{4.1in} 
\\
\vspace*{1cm} Back-door fine-tuning in supersymmetric low scale inflation }
\author{Z.~Lalak$^1$ and  K.~Turzy\'nski$^{1,2}$\\
{\small {}}\\
{\small ${}^1${\it Institute of Theoretical Physics, Warsaw University,}}\\
{\small {\it ul.\ Ho\.za 69, 00-681 Warsaw, Poland;  }}\\
{\small ${}^2${\it Physics Department, University of Michigan,}}\\
{\small {\it 450 Church St., Ann Arbor, MI-48109, USA  }}\\
}
\date{}

\maketitle

\begin{abstract}
Low scale inflation has many virtues  and it has been claimed that its natural 
realisation in  supersymmetric standard model can be achieved rather easily. 
In this letter we have demonstrated that also in this case the dynamics 
of the hidden sector responsible for supersymmetry breakdown and the structure of the soft terms 
affects significantly, and in fact often spoils, the would-be inflationary dynamics. 
Also, we point out that the issue if the cosmological constant cancellation in the 
post-inflationary vacuum strongly affects supersymmetric inflation. It is important to note the crucial difference between
freezing of the modulus and actually stabilising it - the first approach misses parts of the scalar potential which 
turn out to be relevant for inflation.
We argue, that it is more likely that the low scale supersymmetric 
inflation occurs at a critical point at the origin in the 
field space than at an inflection point away from the origin, as the necessary fine-tuning in the second case is typically 
larger. 
\end{abstract}


\section{Introduction}

Implementation of the inflationary paradigm in most 
promising extensions of the Standard Model, inspired by such ideas as supersymmetry or 
extra dimensions, has proven to be a nontrivial and highly demanding task. 
A particular problem of supersymmetric models is related to the existence of the moduli sector, 
consisting of the scalar fields which mix with matter-like scalars and acquire a nontrivial potential 
in the course of supersymmetry breakdown. As pointed out in 
\cite{Ellis:2006ar,Brax:2006ay} 
this complicates inflationary 
dynamics and may easily invalidate the expectations for successful 
inflation based on a single-field approximation
in the presence of frozen moduli. Since this problem has often been overlooked, in the present letter we shall 
discuss it again, in the context of low scale supersymmetric inflation \cite{German:2001tz},
 which clearly exposes 
its significance. 
Among various interesting proposals available in this area, there is the proposal of 
generating inflation by using flat directions of the minimal supersymmetric 
standard model (MSSM) as candidate inflatons \cite{Allahverdi:2006iq}, which we shall use as a specific 
example. 
The necessary curvature of the inflationary potential
arises there due to non-renormalisable corrections to the superpotential and due to the soft terms 
which are borne in the underlying microscopic theory. The problem appears in the form of the 
particular relation between the soft terms, which needs to be fulfilled to a very high accuracy in 
order to produce phenomenologically acceptable inflation. 
The solution put forward in \cite{Enqvist:2007tf} attempts to explain the troublesome relation via the functional 
form of the K\"ahler potential of the modulus controlling the soft terms at the level of the 
microscopic theory -- the four-dimensional supergravity. Also, the form of 
the K\"ahler potential coupling 
of the modulus to the inflaton needs to be prepared accordingly. This would be indeed an elegant 
resolution of the problem, as the choice of the form of the K\"ahler potential and the 
superpotential gives automatically relations between individual couplings in the Lagrangian, 
and supersymmetry could be hoped to control these relations beyond the tree-level expressions. 

Unfortunately, proper embedding of the inflationary version of the MSSM in supergravity  
moderates these expectations. The point is that the modulus sector of the theory does not 
really decouple from the inflationary dynamics. In fact, one needs to minimise and evolve 
the whole system consisting of the inflaton and the modulus at the same footing, and this may easily 
modify the MSSM-level predictions as we 
demonstrate in this letter. Also, we point out that the issue if the cosmological constant cancellation in the 
post-inflationary vacuum affects supersymmetric inflation. It is important to note the crucial difference between
freezing of the modulus and actually stabilising it - the first approach misses parts of the scalar potential which 
turn out to be relevant for inflation. Supergravity corrections and the dynamics of the modulus result in the 
situation, where the single choice of the K\"ahler potential and superpotential becomes replaced by 
the whole series of tunings between Lagrangian couplings and the usual fine-tuning 
problem gets re-created in its full severity.
  
We argue, that it is more likely that the low scale supersymmetric 
inflation occurs at a critical point at the origin in the 
field space than at an inflection point away from the origin, as the necessary fine-tuning in the second case is typically 
larger. 

\section{Small-scale inflation}
\label{sectwo}

\subsection{General remarks}
\label{general}

When one attempts to find an inflationary model in supergravity,
one often finds that the typical scale of the scalar potential
is $V\sim m_{3/2}^2M_P^2$, which for gravitino masses close
to the electroweak scale is about $\sim 10^{-32}M_P^4$.
This is to be compared with the magnitude of the spectrum of primordial
curvature perturbations:
\beq
\mathcal{P}_\mathcal{R} = \frac{1}{24\pi^2\epsilon} \frac{V}{M_P^4}
\label{spec1}
\eeq
and its spectral index:
\beq
n_s = 1-6\epsilon+2\eta \,
\label{spec2}
\eeq
where $\epsilon=(M_P^2/2)(V'/V)^2$ and $\eta=M_P^2(V''/V)$ are the
usual slow-roll parameters. It follows from observations that
$\mathcal{P}_\mathcal{R}\sim 2\times10^{-9}$ while $n_s$ is by a few per cent
smaller than unity \cite{Spergel:2006hy}. 
At first sight, inflation with a low scale
suggested by supergravity appears to fit well into this picture:
one can envision looking for models with very small $\epsilon$ to provide
a correct normalisation of the spectrum
and with a negative $\eta$ is of the order $\mathcal{O}(10^{-2})$ for a correct
value of the spectral index. Therefore, we start our analysis by discussing
to what extent this assumption is realistic in reasonably simple
inflationary models.

When constructing inflationary models, one often looks for the saddle points
of the potential. If such a saddle point exists and
there is only one unstable direction at this point, a possibility of
successful inflation arises.
In the vicinity of a saddle point, we can approximate the inflationary 
potential $V$ as a function of a real scalar field $\varphi\equiv M_P\sigma$ 
corresponding to the unstable direction by:
\beq
V(\varphi) = V_0 \left(1 -\frac{y}{2q}\left(\frac{\varphi}{M_P}\right)^q\right)^2 \, .
\label{inipot}
\eeq
Inflation ends at some field value $\sigma_f$ for which $\epsilon$ becomes
large. Usually this condition is taken as $\epsilon=1$, but it will not
introduce a large error if we identify $\sigma_f$ with the field value for
which $\epsilon$ actually diverges, $\sigma_f^q=2q/y$. The earliest moment
of inflation we shall be interested in is when the characteristic scales
of CMB left the Hubble radius, $N\sim 50$ efolds before the end of inflation.
when the field
value was $\sigma_i$. 
The number of efolds $N$ is given by:
\beq
N = \frac{1}{M_P} \int_{\sigma_f}^{\sigma_i} \frac{V}{V'}\mathrm{d}\sigma
= \frac{1}{4q}\left( \sigma_i^2-\sigma_f^2\right) + \left\{ \begin{array}{ll} \frac{1}{y}\ln\frac{\sigma_f}{\sigma_i} & : q=2 \\ \frac{1}{(q-2)y}\left(\frac{1}{\sigma_i^{q-2}}-\frac{1}{\sigma_f^{q-2}}\right) & : q>2\end{array}\right.
\label{efolds}
\eeq
where $\sigma_i$ corresponds to the value of the scalar field $N$ efolds
before the end of inflation.
In the following, we shall try to estimate crudely $\mathcal{P}_\mathcal{R}$
in terms of the quantities entering directly the inflationary observables,
assuming that inflation takes place close to the saddle point at $\sigma=0$.

If $q=2$ and $1/N\simlt y\ll1$, the second term
dominates
 in (\ref{efolds}) and then $\sigma_i=\sqrt{2q/y}e^{-Ny}$.
One then finds $\eta\approx-y$ and $\epsilon\approx 2ye^{-2Ny}$.
In this regime, $\epsilon$ is bounded from below
and it is impossible to reconcile (\ref{spec1})
with a small scale of the 
potential (if 
the potential (\ref{inipot}) ceases to be an accurate
approximation for $\sigma\sim\sigma_f$ and inflation actually ends at
a smaller value $\sigma_f$ than discussed above, it leads to
smaller $\sigma_i$ and improves
the quality of the rough approximations used above). 
For $y\ll 1/N$, we find that 
$\eta\approx\epsilon\approx 1/(2N)$ and a similar conclusion holds.
We also note that
for $y\simgt1$ the potential (\ref{inipot}) cannot support slow-roll
inflation.

For $q>2$, it follows from (\ref{efolds}) that 
$\sigma_i^{q-2}\approx\frac{1}{N(q-1)y}$. If $Ny\ll1$, one finds
that $\eta=2\epsilon$ which is inconsistent with a low-scale inflation
predicting a realistic scale dependence of the perturbations.
For $Ny\simgt 1$, one gets
\beq
\epsilon \approx \left(N(q-2)\right)^{-(2q-2)/(q-2)}y^{-2/(q-2)} \, ,
\qquad
\eta \approx -\frac{q-1}{q-2}\frac{1}{N} \, .
\eeq
A striking feature of this regime is that $\eta$ has automatically
a realistic value which does not depend on $y$. One can then make
$\epsilon$ arbitrarily small by taking a sufficiently large $y$,
which allows reconciling the predictions (\ref{spec1}) and (\ref{spec2})
of low-scale inflation with observations.

A lesson from this analysis is that single-field
low-scale inflation can be realised
if the potential is sufficiently flat. It is, however, not enough
that only $V'(\varphi_0)$ vanishes at some point $\varphi_0$, since
then the sizes of both slow-roll parameters are similar, contrary to what
is required for a realistic spectral index. If the potential
obeys $V'(\phi_0)=V''(\phi_0)=0$, then one can introduce an arbitrary
hierarchy between the slow-roll parameters and satisfy the observational
constraints.

\subsection{Examples of models of low-scale supersymmetric inflation}

One of the models aiming at lowering the scale of inflation
in supergravity was that of Ref.\ \cite{Adams:1996yd}. It employed
a single chiral superfield with a minimal K\"ahler potential
and the superpotential $W(\phi)=\Delta^{-1}(\phi-M_P)^2$. This choice ensures
that all contributions to terms quadratic in $\phi$ in the potential
cancel out and the model can be effectively described by the potential
(\ref{inipot}) with $q=3$ and $y=12$. Of course, such small value of $y$
did not allow for inflationary scale significantly smaller than the GUT scale.

Another model of small-scale inflation has been proposed 
in Ref.\ \cite{German:2001tz}. The potential used in that analysis
was basically (\ref{inipot}) with an addition of a term
quadratic in $\varphi$. It was shown that satisfactory predictions
for the spectrum of primordial perturbations can be achieved in such
a model for $q>2$ and that the presence of a subleading quadratic term
can quantitatively affect $n_s$, but the normalisation of the power
spectrum is basically left intact. The author also tried to embed
this model in supergravity, in the context of $D$-term inflation.

The recent model of the MSSM inflation \cite{Allahverdi:2006iq}
identifies the inflaton with a flat direction of an unbroken
Minimal Supersymmetric Standard Model (MSSM).  Such a flat direction $\varphi$
can be lifted by a 
nonrenormalisable superpotential $W=\lambda\varphi^n/M_P^{n-3}$
and, in the presence of gravitationally mediated supersymmetry breaking,
its potential generically reads:
\beq
V_\mathrm{MSSM}(\varphi) = \frac{1}{2}m^2\varphi^2 - \frac{\lambda A}{nM_P^{n-3}}\varphi^n +\frac{\lambda^2}{M_P^{2(n-3)}}\varphi^{2(n-1)} \, .
\label{one}
\eeq
Given that the parameters of the potential obey
\beq
A^2 = 8(n-1)m^2 \, ,
\label{two}
\eeq
we have $V'(\varphi_0)=V''(\varphi_0)=0$ at a certain field value $\varphi_0$.
When expanded in terms of $(\varphi-\varphi_0)$, the model can be
well approximated by (\ref{inipot}) around the inflection point.
As demonstrated in the preceding section,
a sufficient number of efolds of inflation may the occur, 
giving rise to primordial density perturbation of
the right magnitude (although in this case $q=3$ which gives 
the spectral index is $n_s=1-4/N$, a bit more that 2$\sigma$ 
off from the WMAP3 measurements). In particular, with the flat directions
$\varphi^3=udd$ or $\varphi=LLe$, we have $n=6$, which allows for the
correct normalization of the spectrum of the curvature perturbations
with suggestive values $m,A\sim\mathcal{O}(\mathrm{1\,TeV})$
of a supersymmetric model solving the hierarchy problem.
It is instructive to compare various energy scales in this setup.
With $A\sim\mathrm{1\,TeV}$ we have $\varphi_0^{n-2}/M_P^{n-2}\sim A/M_P\sim 10^{-16}$, $V(\varphi_0)^{1/4}/M_P\sim (A/M_P)^{1/2}(\varphi_0/M_P)^{1/2}\sim 10^{-10}$, $V'''(\varphi_0)/M_P\sim (A/M_P)^2(\varphi_0/M_P)^{-1}\sim 10^{-28}$ and
$H/M_P\sim (V(\varphi_0)/M_P)^{1/2}\sim 10^{-19}$.
As written in (\ref{one}) and (\ref{two}), the model is severely fine-tuned:
relation (\ref{two}) must be satisfied with the accuracy of 16 digits
to keep the potential sufficiently flat for inflation.
It has, however, been argued that relation (\ref{two}) might
reflect the dynamical properties of the embedding into four-dimension
supergravity with a reasonably motivated ansatz about the K\"ahler 
potential and the superpotential \cite{Enqvist:2007tf}. 

\section{Supergravity embedding}

Although in a realistic model the hidden sector may consist of a plethora
of fields with complicated interactions,
we take for simplicity
only one hidden sector field $T$
and the K\"ahler potential of the form (from now on we set $M_P=1$,
unless specified otherwise): 
\beq
K = \beta\,\mathrm{log}(T+\bar{T})+\sum_{l=1}^\infty Z_{2l}(\phi\bar{\phi})^l \, .
\label{kahass}
\eeq
We further simplify the form of the K\"ahler potential
by taking $Z_2=\kappa/(T+\bar{T})^\alpha$ and $Z_{2l}=0$ for $l>1$.
We also assume that the superpotential is separable, 
$W(T,\phi)=F(T)+G(\phi)$, and we assume that the potential
has an extremum at $(T,\phi)=(T_0,0)$. In addition, we shall
work with the inflaton superpotential which can be written as
$G(\phi)=\phi\tilde{G}(\phi)$, with $\tilde{G}(0)=0$.

\subsection{The cosmological constant problem}

For the rest of this letter we shall relegate the possible D-term contributions to supergravity scalar potential to 
a separate sector of the model which is responsible for cancellation of the cosmological constant, which may contain in 
addition an O'Raifeartaigh $F$-term described below or nonsupersymmetric contributions. The remaining part containing $F$-terms of the inflaton and of the active modulus 
as well as the suitable negative gravitational contribution, we shall simply call (with some abuse of precision) the $F$-term 
potential. 
    
Let us suppose that this $F$-term potential vanishes at the extremum
at $(T_0,0)$. This and the vanishing of $\partial V/\partial T$
at this point allows expressing $F_T$ and $F_{TT}$ in terms of
$W_0\equiv W(T_0,0)$, $T_0$ and $\beta$. In particular, a solution
to $V(T_0,0)=0$ reads:
\beq
F_T=\frac{W_0}{T_0+\bar T_0}\left(e^{\imath\theta}\sqrt{-3\beta}-\beta\right)\, ,
\eeq
where $\theta$ is an unknown angle.
Substituting these results
to the 
matrix of the second derivatives of the scalar potential
$\frac{\partial^2V}{\partial x_i\partial\bar{x}_j}$, where 
$x_i=(T,\bar{T},\phi,\bar{\phi})$, we obtain a block-diagonal form:
\beq
\left. \frac{\partial^2V}{\partial x_i\partial\bar{x}_j} \right|_{(T_0,0)}=
\left(
\begin{array}{cc}
D_1 & 0 \\
0 & D_2
\end{array}
\right) \, ,
\label{secder}
\eeq
where $D_1$ and $D_2$ are $2\times 2$ matrices and, in particular,
\bea
\left.(D_1)_{T\bar T}\right|_{(T_0,0)} &=& -\frac{2(\beta+3)W_0W^\ast_0}{(T_0+\bar T_0)^{2-\beta}} \\
\left.(D_1)_{TT}\right|_{(T_0,0)} &=& -\frac{\sqrt{-3\beta}W_0W^\ast_0}{(T_0+\bar T_0)^{2-\beta}} \left( e^{-\imath\theta}(\beta^2+3\beta+1+\frac{X_T}{\beta})-18e^{\imath\theta}+\frac{9\beta^2+36-2\beta e^{2\imath\theta}}{\sqrt{-3\beta}}\right) \, ,
\eea
where $X_T\equiv F_{TTT}(T_0+\bar T_0)^3/W_0$.
If $\beta\neq-3$ and the superpotential is a nontrivial function of $T$,
it can be shown that
one of the eigenvalues of $D_1$ is negative;
otherwise, the smaller eigenvalue vanishes (for 
$\beta=-3$ and $\alpha=1$,
corresponding to no-scale supergravity, this conclusion is valid
for an arbitrary form of $G$).
There are, therefore, two possibilities:
(i) the hidden sector field $T$ is stabilised, but the point $(T_0,0)$ 
at which the potential vanishes cannot be a local minimum of the potential
(hence, at the true minimum the $F$-term
potential assumes a negative value and we are faced with the usual problem
of the cosmological constant) or (ii) the cosmological constant vanishes,
but the hidden sector field is not stabilised. 
In the following we assume that the case (i) holds and,
somewhat optimistically,
that the cosmological constant
problem is solved by an additional mechanism which
does not affect form of the $\phi$-dependent 
$F$-term contributions to the potential.

To be somewhat more specific let us discuss briefly a supersymmetric uplifting with an additional chiral (O'Raifeartaigh)
 sector in the 
superpotential, which  doesn't depend on the $T$ modulus and fixes the expectation values of its scalar 
components at values much smaller than the Planck scale, $\langle\psi\rangle \ll M_P$, in such a way that the scale of the O'Rafeartaigh $F$-terms in the globally supersymmetric limit is $|F_\psi|^2 \approx \mu^4 \; < \; M_{P}^4$. Then one can expand the supergravity scalar potential in powers of $\psi / M_P$ and leading terms in this expansion 
can be written as, see \cite{Scrucca,Dudas,Westphal},
\begin{equation} \label{or1}
V(T,\psi) = \frac{\mu^4}{8 t^3} + \frac{\Sigma^4}{2} (t - t_0)^2 -\Sigma_0,
\end{equation}
where $t={\rm Re}(T)$, $\Sigma_0 > 0$, and the dependence of the first term on $T$ comes from the factor $e^K$ with 
$K(T) \supset - 3\log(T +  \bar{T})$. With the O'Raifeartaigh sector switched off the scalar potential has a minimum at 
$t=t_0$ with a negative vacuum energy $-\Sigma_0$. However, when one takes $\mu \neq 0$ the minimum becomes shifted to 
$\langle t\rangle = t_0 + \delta$ with $\delta \approx \frac{3 \mu^4}{8 \Sigma^4 t_{0}^{3}}$. Further to this, 
a single tuning 
$\mu^4 \approx 8 t_{0}^3 \Sigma_0$ gives a vanishing vacuum energy density at this minimum. 
So one can uplift, but the relevant question is under which conditions the shift of the minimum due to uplifting, 
and consequently the change of the original inflationary potential, is small. Requesting $\delta \ll t_0$ one obtains 
a condition on $\Sigma, \; \mu$ and $t_0$: $\Sigma > \frac{\mu}{t_0}$ which is equivalent to the condition 
$\Sigma^4 > \frac{3 \Sigma_0}{t_0}$. One can check that if the above condition holds, the gravitino mass before and after 
uplifting changes at the order $\mathcal{O}(\delta)$, hence $m_{3/2}^2$ shifts in fact by a negligible amount. 

In what follows we shall assume the mechanism just described an explicit example of a supersymmetric uplifting, 
which leaves mixed inflationary/modular sector of the model unaffected to any practically required accuracy, which, however, can be achieved at the expense
of further tuning.

\subsection{Supergravity corrections}
\label{sugraco}

For $G(\phi)=(\tilde\lambda/\nu)\phi^\nu$ with $\nu\geq3$
we can expand the scalar potential as:
\beq
V_F(T,\phi) = -V_s(T)+\left[-V_s(T)\left(1+\frac{\alpha}{\beta}\right)+(T_0+\bar{T}_0)^\beta\left( 1+\frac{3\alpha}{\beta}\right)W_0W^\ast_0 \right]\frac{\kappa}{(T+\bar{T})^\alpha} \phi\bar{\phi} 
+\mathcal{O}((\phi\bar{\phi})^{3/2}) \, ,
\label{potbig}
\eeq
where
\beq
-V_s(T) = \frac{(T_0+\bar{T}_0)^2}{-\beta} F^\ast_{\bar{T}}F_T - (T_0+\bar{T}_0)(F_T W^\ast_0+F^\ast_{\bar{T}}W_0)-(\beta+3)W_0W^\ast_0 \, .
\eeq
\subsubsection{Saddle point at $\phi=0$}
\label{subsad}
As argued above, there is a negative
contribution to the vacuum energy from the $F$-term potential (\ref{potbig}), 
which yields a negative contribution
to the mass squared of $\phi$. Without specifying the superpotential $F(T)$ 
it cannot be guaranteed that such a negative contribution
can be compensated by the second term in the square brackets in (\ref{potbig}).
If the direction of $|\phi|$ is unstable, we may try to identify
the requirements necessary for such a saddle point to support
inflation consistent with observations. 
In this case, we denote by $V_0$ the sum of the vacuum energy
at $\phi=0$ and the uplifting needed to set the cosmological
constant to zero at the true minimum of the potential at $\phi\neq0$.
We first proceed to calculating higher
order terms in the expansion in powers of $\phi$, 
i.e.~$V = V_0 + V_2|\phi|^2+V_3|\phi|^3+\ldots$,  
and comparing them to the coefficients of the effective potential
reminiscent of (\ref{one}),
$V=V_0+m^2\varphi^22/2+A\lambda\varphi^\nu/(\nu M_P^{\nu-3})+\lambda^2\varphi^{2(\nu-1)}/M_P^{2(n-3)}$, where
$\varphi/M_P = (\sqrt{\kappa}/(T_0+\bar{T}_0)^{\alpha/2})|\phi|$
is the almost canonically normalised field (as in Ref.\ \cite{Enqvist:2007tf}  we assume that the kinetic term
is $(\partial_\mu\varphi)(\partial^\mu\varphi)$; rescaling to canonical
normalization $\varphi\to\varphi/\sqrt{2}$ can be compensated by 
$\lambda\to2^{(\nu-1)/2}\lambda$, $A\to2^{1/2}A$ and $m^2\to2m^2$, 
which, in particular, does not
change relation (\ref{two})).
At the lowest order in $\kappa$, we then find:
\bea
V_\nu &=& (T_0+\bar{T}_0)^\beta(-1)^k\frac{|\tilde{\lambda}|}{|W_0|}\left[-\left|\frac{1}{\nu}+\frac{\alpha}{\beta}\right|(T_0+\bar{T}_0)(W^\ast_0 F_T+F^\ast_{\bar{T}}W_0)
\label{v6}
+2\left(1-\frac{\beta+3}{\nu}-\alpha\right)W_0W^\ast_0\right] \\
V_{2(\nu-1)} &=& (T_0+\bar{T}_0)^\beta|\tilde{\lambda}|^2\frac{(T_0+\bar{T}_0)^\alpha}{\kappa} \,.
\eea
In (\ref{v6}), $k$ accounts for minimization with respect to 
$\mathrm{arg}(\phi)$ and it is an even (odd) number if the expression in the 
square brackets is negative (positive). The relation between the coupling
constants $\tilde\lambda$ and $\lambda$ is given by
$\lambda = |\tilde\lambda|(T_0+\bar{T}_0)^{\beta+\alpha\nu/2}/\kappa^{\nu/2}$,
and we assume that $\lambda\sim\mathcal{O}(1)$.
Then we can read the coupling $A$ from (\ref{v6}):
\beq
\label{va}
\frac{A}{M_P} = \frac{\nu(-1)^k(T+\bar{T})^{\beta/2}}{|W_0|}\left[-\left|\frac{1}{\nu}+\frac{\alpha}{\beta}\right|(T_0+\bar{T}_0)(W^\ast_0 F_T+F^\ast_{\bar{T}}W_0)+2\left(1-\frac{\beta+3}{\nu}-\alpha\right)W_0W^\ast_0\right] \, .
\eeq
Although the overall form of the scalar potential is somewhat different
from the one discussed in Section \ref{sectwo}, a successful inflation
would be supported mainly by the constant term and the $A$-term, identical
to the one studied there. We can, therefore, follow the analysis
presented in Section \ref{sectwo} in studying conditions
that have to be satisfied for a successful inflation at the saddle point.

Firstly, we need $|m^2|M_P^2/V_0\simlt\mathcal{O}(10^{-2})$ in order not to spoil the
predictions for the spectral index. Secondly, during inflation 
the $A$-term contribution to the potential
must be much larger than the mass term contribution, $1\ll A\lambda \varphi^{n-2}/(n|m^2|M_P^{n-3})$, which gives $|m^2|M_P^2/V_0\ll1/(n(n-1)N)$.
Finally, we need to be in the large $y$ regime of Section \ref{general},
which gives $1\ll AM_P^3/V_0$. With $A^2\sim|m^2|$ this can be satisfied
rather easily.

In addition to various constraints from the tree-level potential,
there may also be potentially dangerous contributions from quantum
corrections, accounted for by $(32\pi^2)^{-1}V''(\varphi)\Lambda^2$
term in the Coleman-Weinberg potential.
A rough condition that the size of these corrections is much smaller than the
size of the tree-level terms can be written as $1\simgt \Lambda^2/\varphi_{i}^2$
(assuming that $(32\pi^2)^{-1}$ gives a satisfactory
suppression).Using the findings of Section \ref{sectwo}, we obtain the condition
$\Lambda^2/M_P^2\simlt (q-2)^2N^2\epsilon$ or, using the normalisation of
the spectrum of the scalar perturbations,
$\Lambda \simlt 10^5 V_0^{1/2}/M_P\gg \sqrt{|m^2|},A$. One expects $\Lambda$ to be of the order of the visible 
supersymmetry breaking scale, that is close to $m$ within a few orders of magnitude. This is consistent with our condition 
if $V_0^{1/2}/M_P$ is significantly larger than $m$. Fortunately, this is indeed the case for the examples discussed in Section \ref{sectwo}. However, one should note that if $V_0 \approx m$, the above consistency condition would imply a very 
large scale of $m$ hence also a high scale of the inflationary Hubble parameter. Thus the condition of negligibility 
of quantum corrections to the inflationary potential implies certain, model dependent, degree of fine-tuning in 
low-scale models.

\begin{figure}
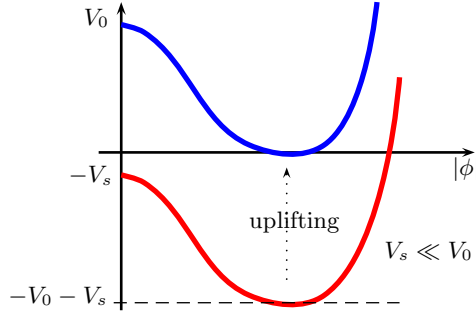

\begin{center}
\psset{unit=1cm}
\pspicture(0.0,0.0)(5,4)
\psline[linewidth=1pt,linecolor=black,linestyle=solid]{->}(0.3,-0.1)(0.3,4)
\psline[linewidth=1pt,linecolor=black,linestyle=solid]{->}(0,2)(5,2)
\put(-0.2,3.7){\small$V_0$}
\put(4.7,1.7){\small$|\phi|$}
\pscurve[linewidth=2pt,linecolor=red,linestyle=solid]{-}(0.3,1.7)(0.6,1.6)(2,0.1)(3,0.1)(4,3)
\pscurve[linewidth=2pt,linecolor=blue,linestyle=solid]{-}(0.3,3.7)(0.6,3.6)(2,2.1)(3,2.1)(3.7,4)
\psline[linewidth=0.5pt,linecolor=black,linestyle=dashed]{-}(0.2,0)(4,0)
\put(-1.2,0.){\small$-V_0-V_s$}
\put(-0.4,1.6){\small$-V_s$}
\put(3.8,0.6){\small$V_s\ll V_0$}
\psline[linewidth=0.7pt,linecolor=black,linestyle=dotted]{->}(2.5,0.2)(2.5,1.8)
\put(2,1){\small uplifting}
\endpspicture
\end{center}
\caption{\em A schematic representation of a shape of the $T=T_0$ section of the potential necessary for successful inflation. \label{figsch}}
\end{figure}

This discussion shows that it is possible to accommodate
a low scale inflation in supergravity given that a number of
conditions are satisfied and a certain degree of fine-tuning arranged for,
as we schematically depict in Figure \ref{figsch}.
Firstly, the supergravity potential must be uplifted
to solve the cosmological constant problem
(the enormous associated fine-tuning should be perhaps treated
as a separate problem);
the uplifting should also leave the $\phi$ dependence of the $F$-term
potential qualitatively unchanged. Secondly, the condition $|m^2|M_P^2 \ll V_0$ tells us, that we need $V_0$ a few orders of magnitude larger than the scale of natural un-tuned vacuum value of the scalar potential, usually negative,  $m_{3/2}^2 M^2$, as 
$m^2 \approx M_{3/2}^2$.  This tells us also that there is a hierarchy between the value
of the scalar potential at the saddle point before uplifting  and its value at the true minimum (un-lifted)
($-V_s$ and $-V_0-V_s$, respectively), which  is needed to alleviate
the $\eta$ problem of supergravity \cite{German:2001tz}
and to provide inflation lasting sufficiently long.
Of course, accommodating these general guidelines in a concrete
model is by itself a challenging task, which we leave for a future 
study.

\subsubsection{Minimum at $\phi=0$: the case of the MSSM inflation}

In addition to the situation discussed in Section \ref{subsad},
there is also an alternative possibility that the mass squared of $|\phi|$
is positive
despite the negative contributions related to the
supersymmetric contributions to the vacuum energy.
Then all formulae presented above are valid also in this case,
with an obvious difference that now $V_0=0$,
and we might try to realise the MSSM inflation in such a setup.
In order to achieve inflection point inflation, we have to
arrange the parameters of the theory in such a way that
the fine-tuning condition (\ref{two}) 
is satisfied.
The relevant relation can be obtained by comparing
(\ref{potbig}) and (\ref{va}).
In the simplest case $W(T)=W_0$, it reduces
to a very simple equation relating two numbers $\alpha$ and $\beta$:
\beq
(3-\beta-6\alpha)^2 = -20(\beta+\alpha+2) \, ,
\eeq
which has rational solutions for $\alpha$ and $\beta$,
as already found in Ref.\ \cite{Enqvist:2007tf}.
However, with $W_T\neq 0$, this simple algebraic relation
between two numbers no longer holds -- the solution for $\alpha$ depends
on the details of the interactions of the hidden sector, namely, the
values of $T_0+\bar{T}_0$ and $F_T/W_0$ at the minimum of the potential.
It is rather difficult to envision a natural origin of the
particular alignment of the
parameters in the K\"ahler potential and the superpotential,
satisfying (\ref{two}) in this more general case 

\subsubsection{Dynamical supergravity corrections}

So far we have discussed 
the supergravity corrections following merely from statics of the theory.
There may be also truly dynamical corrections resulting from slight
changes in the values of the hidden sector fields, as the inflation proceeds.
In order to illustrate this, we shall again consider the case with
$W(T)=W_0$. 
Let us also assume that $\beta\neq-3$ and the cosmological constant
is driven to small values by some unknown nonsupersymmetric mechanism. 
For nonzero values of $\varphi$, we can calculate perturbatively small
shifts $\delta T$ from the value of $T$ at the minimum with $\varphi=0$.
The extremum condition reads then 
$(\partial^2 V(\varphi=0)/\partial T\partial T)\delta T+(\partial V_2/\partial T)\phi\bar{\phi}=0$.
As follows from (\ref{potbig}), a straightforward estimate from the 
supersymmetric part of the potential is that 
$\delta T/(T+\bar{T})\sim\mathcal{O}(\varphi^2/M_P^2)$. 
The contribution from the vacuum energy uplifting sector is unknown,
but it is unlikely to cancel the former one.
Since the canonically normalised
hidden sector field is $\sim\ln (T+\bar{T})$, we can expect that the 
canonically normalised hidden sector field can move slightly during inflation 
but with a velocity much smaller than that of $\varphi$, thereby
giving negligible contributions to actual slow-roll parameters 
(even though the slow-roll
parameters in the directions of the hidden sector fields are generically large).

\subsection{Higher-order corrections}

In Section \ref{sugraco}, we calculated the coefficients $V_0$, $V_\nu$
and $V_{2(\nu-1)}$ of the potential and discussed the viability of
inflation, assuming that these are the most important contributions
to the inflationary potential. Let us now verify this assumption
by calculating higher order corrections.

We start with a calculation of the quartic term in the potential, of
the form $\Delta V(\varphi)=\frac{\Delta\kappa}{8}\phi^4$. 
To make things simple, let us again assume
that only $Z_2$ is nonzero in the expansion of the K\"ahler potential 
(\ref{kahass}) and that $W(T)=W_0$. Then, 
expanding the full scalar potential to the fourth order
in the inflaton field, we obtain:
\beq
\label{v4}
V_4 = e^{\hat{K}}|W_0|^2Z_2^2\left[\frac{1}{2}\hat{K}^T\hat{K}_T - \frac{1}{2}+\hat{K}^T\hat{K}^{\bar{T}}Y_{T\bar{T}}+(\hat{K}^T\hat{K}_T)^{-1}(\hat{K}^T\hat{K}^{\bar{T}}Y_{T\bar{T}})^2\right]\, ,
\eeq
where $Y_{T\bar{T}}=Z_2^{-2}Z_{2T}Z_{2\bar{T}}-Z_2^{-1}Z_{2T\bar{T}}$
and $\hat{K}=K|_{\phi=0}$. In
principle, this coupling is of the same order of magnitude as $A^2$ 
(in units of $M_P$), that is $\sim10^{-32}$, as argued above. Note that the canonically normalised inflaton field is $\varphi=Z_2^{1/2}|\phi|$, hence the powers of $Z_2$ in this expansion are irrelevant and, in particular, $\Delta\kappa=Z_2^{-2}V_4$. 

For the saddle point inflation, the ratio of  the
quartic term (\ref{v4}) to the $A$-term (\ref{v6})
is $\sim(\Delta\kappa\,M_P/A)(\varphi/M_P)^{4-\nu} \sim (A/M_P)^{2/(\nu-2)}$,
so this contribution does not spoil the form of the potential
(\ref{inipot}).

The case of the MSSM inflation is slightly more involved and we start by
estimating the impact of a quartic correction
on the potential (\ref{one}) with relation (\ref{two}). 
For $\Delta\kappa<0$ such a correction will result in shifting the point at 
which $V'=0$ from $\varphi_0$ to $\varphi_0+\Delta\varphi$, where 
$\Delta\varphi$ is given by (at the lowest order in $\Delta\varphi$ and 
$\Delta\kappa$):
\beq
0 = \frac{1}{2}V'''(\varphi_0)\Delta\varphi^2+\frac{1}{2}\Delta\kappa\,\varphi_0^3 \, .
\eeq
Demanding that $1\gg \Delta\varphi/\varphi_0$, we obtain the condition
$1\gg |\Delta\kappa|\varphi_0/V'''(\varphi_0)\sim |\Delta\kappa|\varphi_0^2/A^2 \sim |\Delta\kappa| 10^{-24}$. This is easy to satisfy, as a natural
size for the parameters in the potential is $(A/M_P)^2\sim 10^{-32}$. However,
in the inflationary context we must also make sure that after adding $\Delta V$
the potential remains sufficiently flat. The curvature of the potential
at the new extremum $\varphi_0+\Delta\varphi$ is described by:
\beq
\eta = M_P^2 \frac{V'''(\varphi_0)\Delta\phi}{V(\varphi_0)} \sim \frac{M_P^2}{\varphi_0 A} \sqrt{|\Delta\kappa|} \sim 10^{20} \sqrt{|\Delta\kappa|} \, .
\eeq
From this we infer that $|\Delta\kappa|$ cannot exceed $10^{-40}$ or it will
spoil the flatness of the inflationary potential.
As emphasized
in Ref.\ \cite{Enqvist:2007tf}, the flatness of the inflationary potential can
be preserved if we admit higher order corrections to the K\"ahler potential, 
such as $Z_4(\phi\bar{\phi})^2$. This requires a precise cancellation between
contributions from $Z_2$ and $Z_4$, which can be achieved for $Z_4=\mu Z_2^2$,
where $\mu$ is a number determined from the lower order terms. 
In particular, it follows from the 
considerations above that the parameter $\mu$ must be adjusted at a level of 
$\sim10^{-8}$.
While it is conceivable that both $Z_2$ and $Z_4$ can be simple power
functions of $T+\bar{T}$, the required accuracy of the relative normalization
appears a bit artificial.
Note that ensuring that the scalar potential has the form given by
(\ref{one}) and (\ref{two}) actually requires a number of relations
up to terms of order $\mathcal{O}(\phi^{10})$. The fine-tuning of the
quartic coupling is only one example of these, other fine-tunings must be
made for $Z_6$ to assure that (\ref{v6}) is the only contribution of 
significance, for $Z_8$ to assure that the $\mathcal{O}(\phi^8)$ terms
vanishes etc. As a result, it appears that a single condition (\ref{two})
has been traded for a tower of fine-tunings in the supergravity embedding.

\section{Summary}

Low scale inflation has many virtues  and it has been claimed that its natural 
realisation in  supersymmetric standard model can be achieved rather easily. 
In this letter we have demonstrated that also in this case the dynamics 
of the hidden sector responsible for supersymmetry breakdown and the structure of the soft terms 
may affect significantly, and in fact spoil, the would-be inflationary dynamics. 
Moreover, significant  enhancement is expected , as 
additional hierarchy between the slow-roll parameters has to develop in 
this type of inflationary models. The parameter $\epsilon$ has to be very much 
suppressed due to the COBE normalisation and at the same time the $\eta$ has to be of the order $0.01$ 
to account for the WMAP3 preference for the spectral index smaller than unity. 
And indeed, we have shown that  
a severe back-door fine tuning appears via the necessary arrangements of the hidden sector parameters. 

As the working example we have analysed in some detail the model of 
Ref.\ \cite{Enqvist:2007tf}, whose authors
aim at explaining a severe fine-tuning (\ref{two})
of the parameters of the inflationary potential (\ref{one}) 
by embedding the model
in supergravity so that the soft supersymmetry breaking parameters $A$
and $m^2$ are simple and suggestive functions of the hidden sector field
and the relation (\ref{two}) reflects the structure of the underlying
supergravity model. In this note, we have identified two difficulties
of such a proposal.

The first problem is that Ref.\ \cite{Enqvist:2007tf} 
neglects the dynamics of the hidden sector fields.
If the superpotential of the theory depends on the hidden sector fields, 
the mass of the inflaton may be different than in that analysis
or the configuration
with $\phi=0$ may not even be a minimum of the potential. 
Although this does not preclude the scalar potential from
having an inflection point suitable for inflation, the explanation of the
relation (\ref{two}) by the structure of the underlying supergravity theory
no longer holds in this case.

Secondly, although the relation (\ref{two}) can be neatly
explained at the leading order in the inflaton field by assuming a reasonably
well-motivated form of the K\"ahler potential, higher order corrections can
easily destroy this result unless one fine-tunes the coefficients of the
higher order terms in the K\"ahler potential. Hence, the solution 
of the fine-tuning (\ref{two}) relies
on shifting the fine-tuning to a different sector of the theory.

However, analysis of the above difficulties shows that they should be easier to overcome in models 
where low-scale supersymmetric inflation occurs near a critical point near the origin in the field space. 
In this case the main problem lies in finding a suitable mechanism to cancel the cosmological constant in such a way, that
the  resulting scale of scalar potential at the origin is much higher than $m_{3/2}^2 M_{P}^2$, but the soft breaking 
parameters $m^2$ and $A$ do not have to be precisely tuned to each other, which is the case if the inflection point needs to be 
created away from the origin. One may also hope that the origin, that is the point of enhanced symmetry, is more likely place 
to find naturally small derivatives needed for successful inflation. 
Another lesson which may be drawn from our discussion is that the hidden sector with more than a single active modulus 
controlling the soft terms and vacuum energy would be more suitable to fulfill all the  constraints, in particular to create
 a supersymmetric minimum with vanishing cosmological constant.

\section*{Acknowledgments}

\vspace*{.5cm}
\noindent We thank S.~Pokorski for discussions.
This work was partially supported by the EC 6th Framework
Programme MRTN-CT-2006-035863,  by the EC 6th Framework
Programme MRTN-CT-2004-503369, by the  grant MNiSW  N202 176 31/3844
and  by TOK Project  MTKD-CT-2005-029466.
Z.L.~thanks CERN Theory Division for hospitality. K.T.~is supported by the
US Department of Energy.


\vspace*{.5cm} 

\begin{thebibliography}{99}
\bibitem{Ellis:2006ar}
  J.~R.~Ellis, Z.~Lalak, S.~Pokorski and K.~Turzynski,
  JCAP {\bf 0610} (2006) 005
  [arXiv:hep-th/0606133].

\bibitem{Brax:2006ay}
  P.~Brax, C.~van de Bruck, A.~C.~Davis and S.~C.~Davis,
  JCAP {\bf 0609} (2006) 012
  [arXiv:hep-th/0606140].

\bibitem{German:2001tz}
G.~German, G.~G.~Ross and S.~Sarkar,
  Phys.\ Lett.\  B {\bf 469} (1999) 46
  [arXiv:hep-ph/9908380];
  G.~German, G.~G.~Ross and S.~Sarkar,
  Nucl.\ Phys.\  B {\bf 608} (2001) 423
  [arXiv:hep-ph/0103243].

\bibitem{Allahverdi:2006iq}
  R.~Allahverdi, K.~Enqvist, J.~Garcia-Bellido and A.~Mazumdar,
  Phys.\ Rev.\ Lett.\  {\bf 97} (2006) 191304
  [arXiv:hep-ph/0605035].

\bibitem{Enqvist:2007tf}
  K.~Enqvist, L.~Mether and S.~Nurmi,
  arXiv:0706.2355 [hep-th].

\bibitem{Spergel:2006hy}
  D.~N.~Spergel {\it et al.}  [WMAP Collaboration],
  Astrophys.\ J.\ Suppl.\  {\bf 170} (2007) 377
  [arXiv:astro-ph/0603449].

\bibitem{Adams:1996yd}
  J.~A.~Adams, G.~G.~Ross and S.~Sarkar,
  Phys.\ Lett.\  B {\bf 391} (1997) 271
  [arXiv:hep-ph/9608336].


\bibitem{Randall:1994fr}
  L.~Randall and S.~D.~Thomas,
  Nucl.\ Phys.\  B {\bf 449} (1995) 229
  [arXiv:hep-ph/9407248].

\bibitem{Randall:1995dj}
  L.~Randall, M.~Soljacic and A.~H.~Guth,
  Nucl.\ Phys.\  B {\bf 472} (1996) 377
  [arXiv:hep-ph/9512439].

\bibitem{Scrucca}
  M.~Gomez-Reino and C.~A.~Scrucca,
  JHEP {\bf 0605} (2006) 015
  [arXiv:hep-th/0602246].

\bibitem{Dudas}
  E.~Dudas, C.~Papineau and S.~Pokorski,
  JHEP {\bf 0702} (2007) 028
  [arXiv:hep-th/0610297].

\bibitem{Westphal}
  M.~Serone and A.~Westphal,
  arXiv:0707.0497 [hep-th].

\end{thebibliography}
\end{document}